%% file: 00_main.tex
\documentclass[sigconf]{acmart} 
\AtBeginDocument{%
  \providecommand\BibTeX{{%
    \normalfont B\kern-0.5em{\scshape i\kern-0.25em b}\kern-0.8em\TeX}}}

\usepackage{xspace}
\newcommand{\system}{SPIRE\xspace} 

\usepackage{array}
\usepackage{newunicodechar}
\newunicodechar{ː}{\textlengthmark}
\newcolumntype{L}[1]{>{\raggedright\arraybackslash}p{#1}}
\usepackage[T1]{fontenc}
\usepackage[utf8]{inputenc}
\usepackage{textcomp}
\usepackage{newtxtext,newtxmath} 
\usepackage{tipa}    
\usepackage{balance}
\usepackage{colortbl}   
\definecolor{S1Color}{RGB}{217,217,217} 
\definecolor{S2Color}{RGB}{221,212,246} 

\newcommand{\tagbox}[2]{%
    {\setlength{\fboxsep}{3pt}\colorbox{#1}{#2}}%
}

\copyrightyear{2026}
\acmYear{2026}
\setcopyright{cc}
\setcctype{by}
\acmConference[IUI '26]{31st International Conference on Intelligent User Interfaces}{March 23--26, 2026}{Paphos, Cyprus}
\acmBooktitle{31st International Conference on Intelligent User Interfaces (IUI '26), March 23--26, 2026, Paphos, Cyprus}
\acmPrice{}
\acmDOI{10.1145/3742413.3789137}
\acmISBN{979-8-4007-1984-4/2026/03}

\begin{document}

\title[Pedagogical Program Synthesis for Interactive Learning]{Teaching Spell Checkers to Teach: Pedagogical Program Synthesis for Interactive Learning}

\author{Momin N. Siddiqui}
\orcid{0000-0003-1874-7789}
\affiliation{%
  \institution{Georgia Institute of Technology}
  \city{Atlanta}
  \state{Georgia}
  \country{USA}
}
\email{momins@gatech.edu}

 \author{Vincent Cavez}
\orcid{0000-0002-6662-7362}
\affiliation{%
  \institution{Stanford University}
  \city{Stanford}
  \state{California}
  \country{USA}
}
\email{vcavez@stanford.edu}

 \author{Sahana Rangasrinivasan}
 \orcid{0000-0003-1210-0124}
 \affiliation{%
  \institution{University of Buffalo}
  \city{Buffalo}
  \state{New York}
   \country{USA}
 }
 \email{srangasr@buffalo.edu}

\author{Abbie Olszewski}
\orcid{0000-0001-6220-9676}
 \affiliation{%
  \institution{University of Nevada}
  \city{Reno}
  \state{Nevada}
   \country{USA}
 }
\email{aolszewski@unr.edu}

\author{Srirangaraj Setlur}
\orcid{0000-0002-7118-9280}
 \affiliation{%
  \institution{University at Buffalo}
  \city{Buffalo}
  \state{New York}
   \country{USA}
 }
 \email{setlur@buffalo.edu}

\author{Maneesh Agrawala}
\orcid{0000-0002-8996-7327}
\affiliation{%
  \institution{Stanford University}
  \city{Stanford}
  \state{California}
  \country{USA}
}
 \email{magrawala@stanford.edu}

\author{Hari Subramonyam}
\orcid{0000-0002-3450-0447}
\affiliation{%
  \institution{Stanford University}
  \city{Stanford}
  \state{California}
  \country{USA}
}
 \email{harihars@stanford.edu}

\renewcommand{\shortauthors}{Siddiqui, et al.}

\begin{abstract}
Spelling taught through memorization often fails many learners, particularly children with language-based learning disorders who struggle with the phonological skills necessary to spell words accurately. Educators such as speech-language pathologists (SLPs) address this instructional gap by using an inquiry-based approach to teach spelling that targets the phonology, morphology, meaning, and etymology of words. Yet, these strategies rarely appear in everyday writing tools, which simply detect and autocorrect errors. We introduce \system (Spelling Inquiry Engine), a spell check system that brings this inquiry-based pedagogy into the act of composition. \system implements Pedagogical Program Synthesis, a novel approach for operationalizing the inherently dynamic pedagogy of spelling instruction. \system represents SLP instructional moves in a domain-specific language, synthesizes tailored programs in real-time from learner errors, and renders them as interactive interfaces for inquiry-based interventions. With \system, spelling errors become opportunities to explore word meanings, word structures, morphological families, word origins, and grapheme-phoneme correspondences, supporting metalinguistic reasoning alongside correction. Evaluation with SLPs and learners shows alignment with professional practice and potential for integration into writing workflows.
\end{abstract}


\begin{teaserfigure}
  \includegraphics[width=\textwidth]{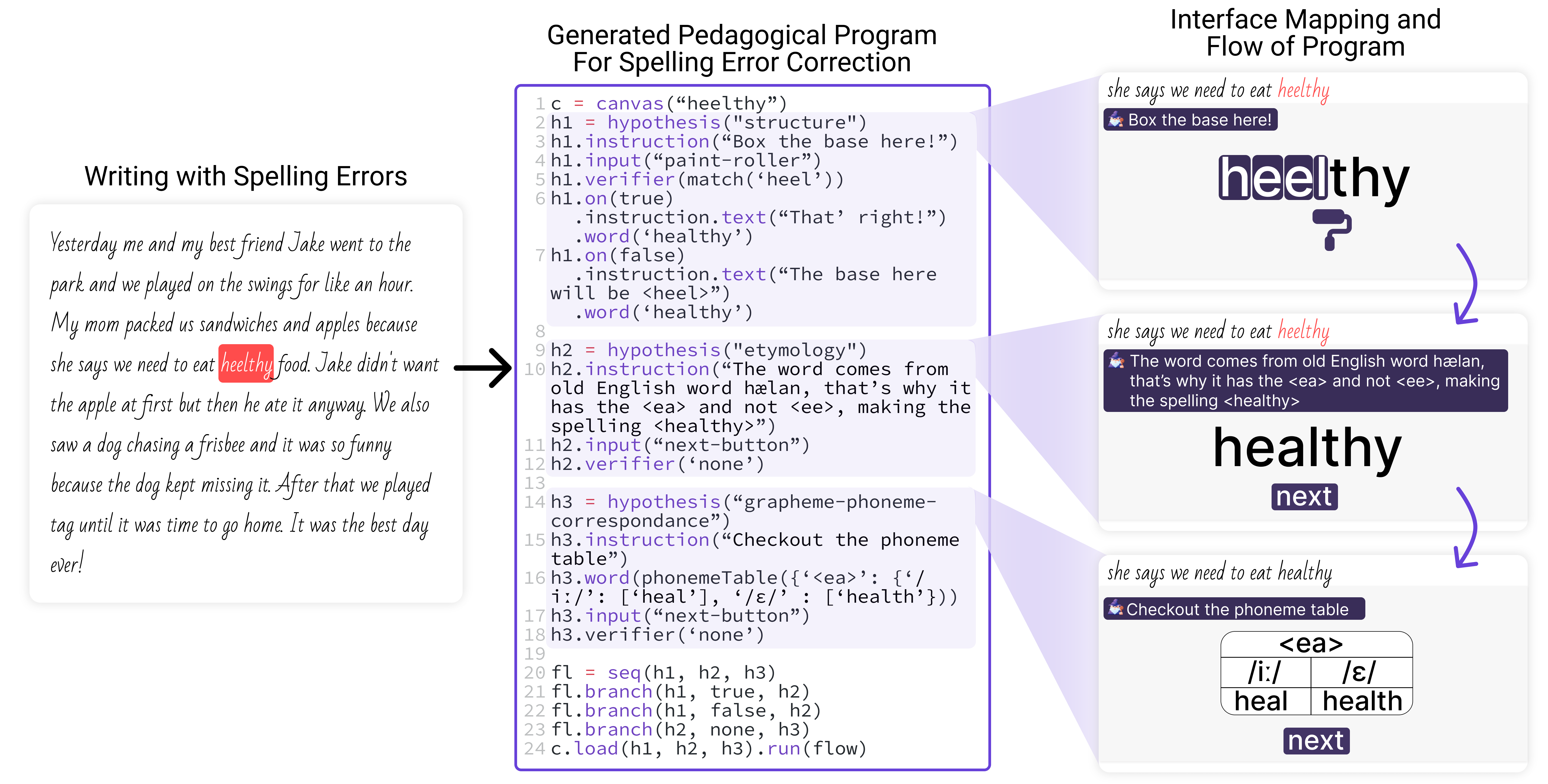}
  \caption{This figures illustrates how \system identifies misspellings (left), generates a program for intervention (center) and subsequently renders them in the interface (right) for learner's inquiry to surface word related insights}
  \label{fig:teaser}
\end{teaserfigure}

\maketitle

\input{01_intro}
\input{02_relatedwork}
\input{03_ux}

\input{04_DSL}
\input{05_system}
\input{06_tech_evaluation}
\input{07_user_evaluation}
\input{08_discussion}

\input{09_conclusion}
\balance{}
\bibliographystyle{ACM-Reference-Format}
\bibliography{99_refs}

\end{document}

%% file: 01_intro.tex
\section{Introduction}

School-age students often learn spelling by memorization~\cite{hegland2021beneath}. However, the way in which a word is spelled is more than just a convention of letters. It reflects patterns of meaning, sound, structure, and spelling conventions that connect words to one another and to their histories. For instance, consider the word `reach' meaning to extend or stretch out. A majority of us can spell this word with ease, and are unlikely to spell it as say <reech>, in which the long /i/ sound is mapped to the grapheme <ee>. In fact, English represents the /i/ sound in many ways, including \textit{<ee>} as in see, \textit{<ea>} as in reach, \textit{<ie>} as in piece, and \textit{<e-e>} as in these. Memorization helps us recall the correct form. However, for many learners, especially children with \textit{language-based learning disorders}, rote memorization is insufficient because it recruits only one aspect of the lexical quality of the word (phonology or orthography/spelling)~\cite{perfetti2007reading,seidenberg1989distributed}. What supports their learning is understanding the logic behind these alternatives; for example, noticing that `reach' and `teach' both use the <ea> to represent \textit{/iː/}, and that this pattern traces back to the Old English \textit{reccan}~\cite{etymonline_reach}. Seen this way, spelling becomes less an arbitrary code to memorize and more a \textit{system of patterns} that can be reasoned about, making the language more learnable and meaningful~\cite{tbox2}. 
In fact, learners who struggle with spelling through memorization are often referred to speech and language therapy. In therapy, speech-language pathologists (SLPs) use \textit{inquiry} as an approach to help learners make sense of how words are spelled and built~\cite{bowers2006teaching}. For example, if a learner writes <reech> for ``reach'', the SLP may begin by asking how the learner decided on that spelling. If the learner explains that /i/ is often spelled <ee> as in ``see'' or ``speech'', the SLP can validate this reasoning, then introduce a contrast, such as ``teach'' or ``teacher'', to highlight the alternative <ea> spelling. With a different error, such as using <sine> for ``sign'', the SLP may draw attention to the silent <g> and connect it to morphologically related words like ``signal'' or ``signature''. This practice, known as \textit{Structured Word Inquiry} (SWI)~\cite{effects_bowers_2010}, turns spelling into a process of investigation rather than memorization, drawing on phonology (the relationship between sounds and letters), semantics (meaning of the word), morphology (the structure and meaning of word parts such as roots, prefixes, and suffixes), and etymology (the historical origins and evolution of words) to explain why words are spelled the way they are. Through these inquiry-based interventions, SLPs help learners build \textit{durable} knowledge that extends beyond a single correction to a broader understanding of how written language works. Consequently, they learn how to independently investigate words.

In contrast, everyday writing tools take a much narrower approach to spelling errors. A misspelling is detected, underlined, and replaced with the most probable correction. Whether a learner types \textit{``reech''} because they overgeneralized a phonics rule, misapplied a morphological pattern, or simply hit the wrong key, the system's response is the same. This transactional design supports error-free text, but it bypasses the chance to engage learners in noticing patterns or testing hypotheses. This is a missed opportunity because therapy time is scarce~\cite{exploring_lewis_2025} and spelling occurs within these tools in the context of authentic writing tasks, such as science reports and essay writing. Without support in these everyday contexts, the carryover of SWI pedagogy from therapy to classroom writing is limited. What is needed is a way to embed the principles of SWI into the tools learners already use, turning each correction into a brief, \textit{situated learning experience}. This motivates our central research question: \textit{How might we bring the rich, inquiry-based pedagogy of SLPs into everyday writing tools such that spelling errors become an opportunity to learn about the writing system? }

In this work, we introduce \system (Spelling Inquiry Engine), which brings structured word inquiry into everyday writing by synthesizing short, just-in-time instructional sequences from each error. Central to our approach is \textit{pedagogical program synthesis}, a form of probabilistic program synthesis tailored to instruction: given a misspelling and writing context such as surrounding text, 
\system infers a distribution over likely causes (e.g., grapheme-phoneme mapping, morphological structure, etc.) and then composes a bounded program consisting of three to five steps based on established SLP interventions represented as a domain-specific language (DSL). Each program is designed to guide the learner through a brief cycle of interactive inquiry and hypothesis testing before fading back into the flow of writing. Programs are optimized for just the necessary interventions and branch adaptively on the learner's next action. This design addresses two core challenges: the system only ever sees the misspelling, not the learner's whole reasoning, and effective instruction must change depending on how the learner responds. At runtime, \system compiles the synthesized program into a compact inline panel for inquiry. 

Similar to the <Healthy> inquiry in Figure~\ref{fig:teaser}, the program may ask the learner, through a short interactive speech prompt, to explain how they arrived at a spelling like <reech>. If the learner responds that /i/ is often spelled -<ee> as in ``see'' or ``speech,'' the program can validate this reasoning and then, using simple visual widgets, introduce a contrast such as ``teach'' or ``teacher'' to highlight the <ea> pattern. It may also display the morphological family of reach (e.g., reaching, reachable, reached) or visualize its Old English root \textit{reccan}, showing how etymology preserves the <ea> spelling. By combining spoken dialogue with lightweight visual widgets, the system mirrors how SLPs use multiple modalities in practice, including talking through different student hypotheses while also drawing or writing on paper, thereby keeping the interaction brief but engaging. We evaluate \system through expert appraisal with SLPs to examine whether synthesized programs align with professional practice using 25 authentic misspellings drawn from ecologically valid contexts. We also conducted a preliminary user study with learners to assess the usability of rich interventions within everyday writing tools. 

Our contributions include: (1) a Pedagogical Program Synthesis that represents SLP's instructional moves in a domain-specific language and assembles conditional programs from learner errors in real-time, (2) \system, a spell-checker that embeds the synthesized programs into lightweight, interactive interfaces, and (3) an empirical evaluation with experts and learners demonstrating how \system aligns with professional practice and integrates into writing workflows. 

%% file: 02_relatedwork.tex
\section{Related Work}
Our work builds on three interconnected bodies of research: spelling development and learning difficulties in school-age students, evidence-based interventions designed to support struggling spellers, and recent efforts leveraging AI to deliver adaptive literacy instruction.\newline

Vellutino et al. \cite{specific_vellutino_2004} synthesized four decades of empirical work, establishing that difficulties in phonological awareness, rapid naming, and verbal working memory constitute core deficits that persist in development. Longitudinal studies of upper elementary students with reading or spelling disabilities reveal how these difficulties in phonological information processing evolve throughout formal schooling \cite{developmental_schmidt_2021}. However, the presentation of symptoms varies considerably between individuals, requiring adaptive intervention approaches \cite{developmental_kuerten_2019}.
Corpus-based documentation allowed for an empirical understanding of characteristic dyslexic error types. DysList systematically cataloged error patterns in multiple languages \cite{dyslist_rello_2014}, while subsequent analysis identified specific phoneme-grapheme correspondence failures that persist despite repeated exposure to high-frequency vocabulary \cite{analysis_cidrim_2024}. These difficulties extend beyond traditional literacy contexts into digital environments, where searchers with dyslexia exhibit distinct information-seeking behaviors requiring specialized interface considerations \cite{understanding_morris_2018}.

Ehri's phase model \cite{phases_ehri_1995, learning_ehri_2005} describes progression through distinct developmental stages: pre-alphabetic, partial alphabetic, full alphabetic, and consolidated alphabetic, each characterized by qualitatively different approaches to word recognition. This framework establishes grapheme-phoneme knowledge as the basis for reading acquisition in alphabetic languages \cite{graphemephoneme_ehri_1998}. However, Silliman et al. \cite{spelling_silliman_2006} demonstrated that preadolescents with atypical language skills show distinct error patterns reflecting deficits in phonological, morphological, and orthographic dimensions, indicating that effective intervention must address multiple representational systems simultaneously.

Müller et al. \cite{syllablebased_mller_2020} found that syllable-based reading instruction produces significant improvements in both word-reading accuracy and comprehension in second-grade students. 

Meta-analytical findings demonstrate more severe reading impairments in deep orthographies such as English compared to shallow orthographies with consistent grapheme-phoneme mappings \cite{orthographic_carioti_2021}, suggesting that intervention approaches must account for the characteristics of the writing system. This has led to challenges to phonics-centric instruction, with arguments that English orthography encodes morphological and etymological information often obscured by phonics alone, advocating instead for teaching the logic of the spelling system \cite{beyond_bowers_2017}. However, structured phonics approaches using word boxes and word sort procedures have been proven effective in facilitating both word recognition and spelling development \cite{facilitating_joseph_2002}, indicating that systematic instruction remains valuable even as the field recognizes the limitations of purely phonological approaches.

Beyond specific intervention techniques, broader pedagogical frameworks shape how spelling instruction is delivered and how learners engage with orthographic knowledge. Inquiry-based learning positions learners as active constructors of knowledge through systematic investigation rather than passive recipients of instruction. This approach follows five distinct phases that characterize the inquiry cycle: orientation, conceptualization, investigation, conclusion, and discussion \cite{phases_pedaste_2015}. Theoretical foundations align with the contemporary understanding of active knowledge construction \cite{inquirybased_friesen_2013}, although empirical evidence reveals essential nuances. Meta-analytical findings show moderate positive effects on learning outcomes, with instructional guidance playing a critical role in determining effectiveness \cite{metaanalysis_lazonder_2016}. Unguided discovery learning proved to be less effective than structured inquiry with appropriate scaffolding, indicating that learner autonomy must be balanced with systematic support. Application to language pedagogy shows that inquiry-based teaching develops a deeper linguistic understanding compared to traditional instruction \cite{inquirybased_lee_2014}, but substantial barriers to implementation persist. Institutional constraints, time limitations, and teachers' resistance to constructivist approaches create friction in classroom contexts \cite{challenges_dorier_2013}. These challenges are amplified for learners with spelling difficulties who need additional scaffolding to balance open-ended exploration with systematic investigation of orthographic principles.

\subsection{Intervention Methods}

Intervention approaches for struggling spellers have predominantly focused on phonological processing and phonics-based instruction, building on the established relationship between phonological awareness and reading acquisition. Intensive remedial instruction targeting phonological processing can prevent reading failure in young children \cite{preventing_torgesen_1999, intensive_torgesen_2001}, with explicit systematic phonics instruction producing better outcomes for at-risk children compared to implicit methods \cite{role_foorman_1998}. Early literacy support programs that combine phonological training with reading practice have shown effectiveness \cite{evidence_hatcher_2006}, and systematic reviews confirm that phonics training benefits English-speaking poor readers, although the effect sizes remain modest \cite{phonics_mcarthur_2018}. Individual response patterns vary considerably, and long-term outcomes depend heavily on the intensity and duration of the intervention.
Multisensory instructional approaches have shown particular promise, with significant improvements documented when visual, auditory, and kinesthetic modalities are integrated \cite{impact_schlesinger_2017}. Color-coded onset-rime decoding interventions have proven effective for first-grade students at serious risk for reading disabilities \cite{effectiveness_hines_2009}. However, critical reviews note that despite advances in understanding dyslexia, the effectiveness of treatment remains variable, with many struggling readers requiring sustained and intensive intervention \cite{current_alexander_2004}. Evidence-based interventions must create a "virtuous circle" where gains in one area support the development of related skills \cite{evidencebased_snowling_2011}.

Structured Word Inquiry offers a fundamentally different instructional approach that recognizes English spelling as a morphophonemic system where spellings represent sound, meaning, and history in an orderly manner. Rather than treating English orthography as a collection of exceptions and irregular rules, SWI demonstrates that the spelling system becomes highly consistent when words are contextualized within their morphology, phonology, etymology, and definition. Bowers and Kirby \cite{effects_bowers_2010} found that morphological instruction produces significant effects on vocabulary acquisition, establishing that teaching morphemic structure supports both spelling and comprehension. This approach engages children in generating and testing hypotheses about how the spelling system works, helping them understand why words are spelled the way they are, an investigative process that aligns with inquiry-based learning principles discussed previously. Colenbrander et al. \cite{assessing_colenbrander_2021} conducted a randomized controlled trial assessing the effectiveness of SWI for students in grades 3 and 5 with reading and spelling difficulties, demonstrating improved performance on both trained and untrained words.

In practice, SWI instruction is typically delivered by speech-language pathologists working with small groups of students, providing the individualized attention and expert guidance necessary for effective implementation. However, the growing demand for specialized literacy support has created significant shortages of qualified SLPs \cite{addressing_squires_2013}, limiting access to these evidence-based interventions. Colenbrander et al.'s \cite{assessing_colenbrander_2021} findings revealed that SWI instruction delivered by teaching assistants with limited training produces significantly less effective outcomes compared to expert-delivered instruction, highlighting the importance of pedagogical fidelity in any delivery method. This expertise gap motivates investigation of whether computational systems can faithfully operationalize the reasoning processes that make expert SWI instruction effective, a prerequisite for eventual scalability that we address through expert evaluation of pedagogical validity.

\subsection{Digital Tools for Teaching Language}

Technological interventions for spelling and vocabulary instruction have evolved along several trajectories, each addressing different aspects of literacy support for children with dyslexia. One line of work has explored tangible interfaces to reduce cognitive load through concrete representations of abstract linguistic concepts. These systems demonstrated that physical manipulatives support young language learners through embodied learning experiences \cite{design_fan_2018, spellbound_pandey_2011}. This approach has been adapted for Chinese learners \cite{character_fan_2019}, addressing the distinct challenges of character-based writing systems where visual complexity and stroke order play central roles.

Adaptive spelling correction systems represent a second trajectory, accommodating characteristic dyslexic error patterns through specialized spellcheckers trained on dyslexic error corpora \cite{spellchecker_rello_2015, computerbased_rello_2015}. These systems learn individual error patterns to provide personalized correction suggestions \cite{polispell_li_2013}, representing significant advances in accommodating dyslexic spelling patterns. However, they remain fundamentally corrective rather than instructional, helping users spell correctly without developing a deeper orthographic understanding.

Gamification has emerged as a third approach, using game mechanics to maintain engagement during spelling practice \cite{gamification_gooch_2016}. Game-based systems have been developed to address spelling difficulties in different linguistic contexts \cite{game_rauschenberger_2015, lexilearn_weerasinghe_2025, gamification_dymora_2019}. Comprehensive reviews of these technologies note that most remain limited to phonetic-based approaches \cite{technologies_rauschenberger_2019}, failing to capture the comprehensive inquiry-based methodology that characterizes effective SWI instruction. Context-aware writing assistance has shown promise in improving outcomes \cite{design_wu_2019}, yet even sophisticated adaptive systems lack frameworks to support hypothesis generation and testing that make SWI effective.

\subsection{AI Pedagogical Techniques}

AI-assisted literacy support has evolved from error correction systems \cite{contextual_gamon_2008} to more sophisticated instructional approaches that adapt to individual learners. The emergence of large language models has particularly enabled generative methods that create personalized learning experiences.
Goodman et al. \cite{lampost_goodman_2022} designed LaMPost, an AI-assisted writing tool for adults with dyslexia that helps compose emails while preserving user agency. Wang et al. \cite{10.1145/3038535.3038542} developed conceptual motivation models specifically for students with dyslexia, recognizing that practical assistive learning tools must take into account motivational factors alongside functional capabilities. 
Leong et al. \cite{putting_leong_2024} demonstrated that AI-enabled context personalization for vocabulary learning improves motivation by situating words within personally relevant scenarios. Attygalle et al. \cite{10.1145/3708359.3712073} explored text-to-image generation for vocabulary learning using the keyword method, finding that externalizing memorable visual links significantly increases memory retention. Peng et al. \cite{storyfier_peng_2023} developed Storyfier, exploring vocabulary support through narrative contexts generated by text models. Wang et al. \cite{learnmate_wang_2025} extended this with LearnMate, providing personalized learning plans and adaptive support. Investigation of AI-based support in speech-language pathology has documented both opportunities and concerns about bias and clinical validity \cite{exploring_lewis_2025}, with particular attention to providing equitable care for culturally and linguistically diverse children. Design explorations of AI-supported home practice in speech therapy reveal how varying levels of AI oversight can provide informational, emotional, and practical support to parents implementing therapy techniques \cite{10.1145/3706598.3713986}, although questions remain about alignment with professional therapeutic goals.

Conversational and Socratic approaches have emerged for fostering critical thinking through interactive dialogue. Ruan et al. \cite{10.1145/3397481.3450648} developed EnglishBot, an AI-powered conversational system for second language learning that adapts feedback based on student interaction, demonstrating greater engagement compared to traditional listen-and-repeat interfaces. Systems using AI to probe student understanding through targeted questioning enable scalable oral assessment \cite{socratic_hung_2024}, while question-driven exploration for personalized education \cite{socratiq_jabbour_2025} and investigations of whether AI partners can empower learners to ask critical questions \cite{ai_maiti_2025} suggest that appropriately designed systems scaffold metacognitive skills. Metacognitive tutoring agents have proven to be effective in teaching inquiry-driven modeling in science education, demonstrating that intelligent tutoring systems can scaffold authentic scientific inquiry processes \cite{10.1145/2678025.2701398}. However, these conversational approaches, while effective for verbal reasoning, may not optimally support the hands-on, constructive activities central to structured word inquiry.

Program synthesis offers an approach particularly well suited to inquiry-based learning contexts requiring structured artifacts. Unlike conversational AI that responds to queries, program synthesis can generate interactive exercises and examples tailored to specific learning objectives. Theoretical foundations for program synthesis \cite{program_gulwani_2017} have been extended by demonstrations that large language models effectively perform synthesis tasks \cite{program_austin_2021}, including interactive synthesis by augmented examples \cite{interactive_zhang_2020}. Recent work has explored how hypothesis search enables inductive reasoning with language models \cite{wang2024hypothesis}, providing mechanisms for iterative refinement of generated artifacts. Synthesis and verification pipelines demonstrate how LLM-generated outputs can be iteratively corrected through automated verification and feedback loops \cite{10.1145/3731209}, achieving high success rates through repeated refinement cycles. Kazemitabaar et al. \cite{codeaid_kazemitabaar_2024} evaluated CodeAid in classroom deployment, documenting how LLM-based synthesis balances exploration with scaffolding. For inquiry-based spelling instruction, program synthesis enables dynamic generation of morphological investigations adapted to learners' current hypotheses, providing the structured yet flexible framework necessary for effective SWI implementation at scale.

%% file: 03_ux.tex
\section{User Experience}

\begin{figure*}
    \centering
    \includegraphics[width=1\textwidth]{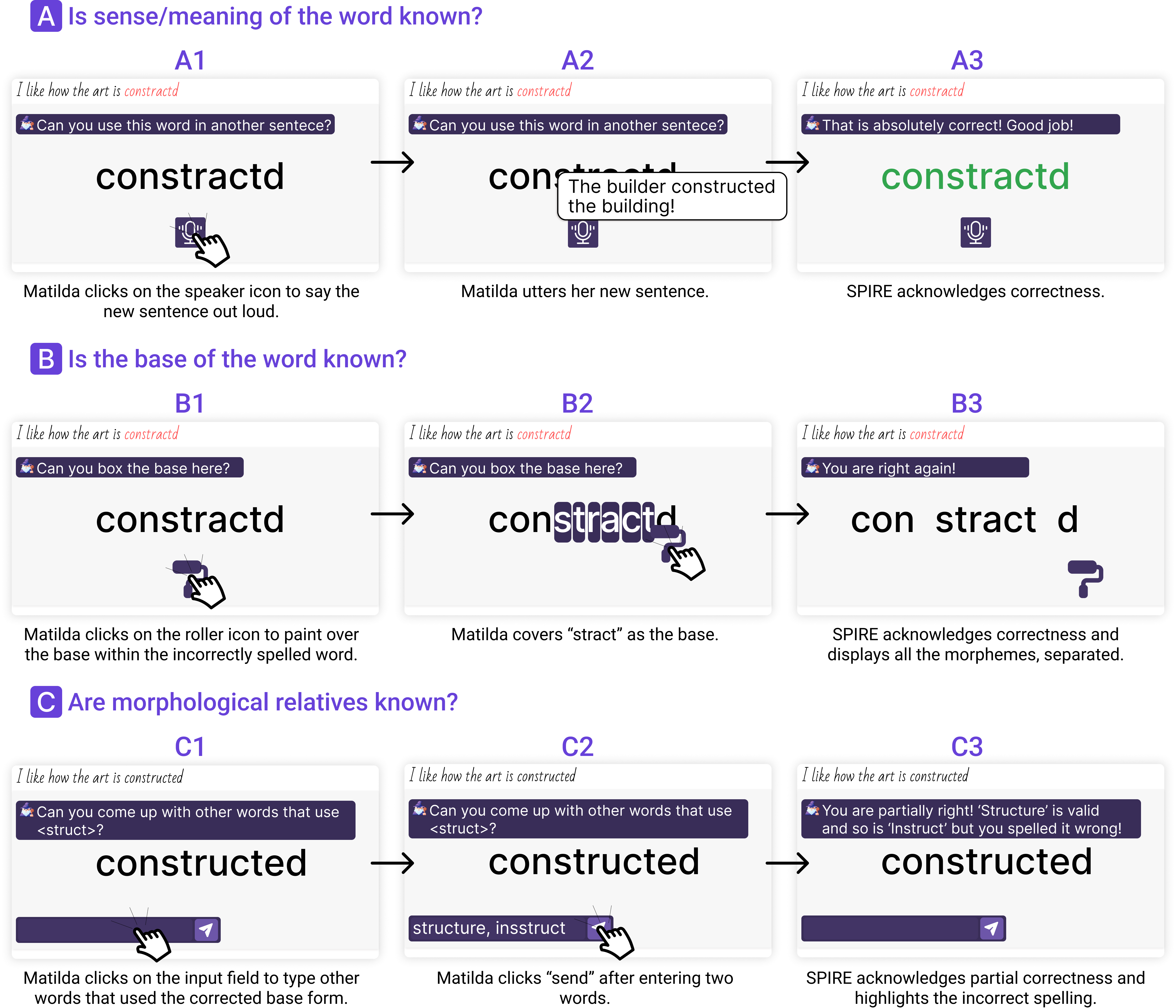}
    \caption{This figure demonstrates the usage scenario. We showcase the learner going through three interventions for the following inquiries: (A) ``Is the sense/meaning of the word known?'', (B) ``Is the base of the word known?'' and (C) ``Are the morphological relatives known?''}
    \label{fig:ux}
\end{figure*}

At a high level, \system transforms traditional spell checking into an interactive process of linguistic inquiry. When a learner misspells a word, \system invites them to explore its structure, meaning, and sound rather than simply replacing it with a correction. Each spelling mistake becomes an opportunity to reason about English orthography through short, guided activities embedded into the writing interface.

\subsection{Usage Scenario}
Matilda opens her word processor, which has \system integrated into it. As she types, she writes: \textit{``I like how the art of constractd.''} The system flags \textit{constractd} as a possible error by enclosing it in a red rectangle. When Matilda clicks on the word, the red highlight transitions into a red text indicator, and an inquiry canvas opens below her writing (Fig.~\ref{fig:ux}A1).

The system begins by asking Matilda to use the word in a sentence---a diagnostic step to determine whether she understands its meaning. She clicks on a microphone icon (Fig.~\ref{fig:ux}A1) and says aloud: \textit{``The builder constructed the building.''} The system transcribes her response and verifies that her usage reflects an appropriate understanding of \textit{construct}. Since her sense of meaning is correct, it affirms her answer (Fig.~\ref{fig:ux}A3) and proceeds to the next inquiry.

Next, \system invites Matilda to \textit{``box the base``} of the word (Fig.~\ref{fig:ux}B1), a structured word inquiry practice where learners identify the morphemic base. Using a paint roller tool, Matilda drags across the portion she believes to be the base and highlights \textit{stract} (Fig.~\ref{fig:ux}B2). Although not fully correct, this action shows that she recognizes the structural pattern embedded within \textit{construct}. The system records this as partial evidence of her morphological awareness and proceeds to test a new hypothesis related to graphemes.

At this point, the word is corrected to \textit{constructed}, revealing how \textit{<u>} replaces \textit{<a>} and an \textit{<e>} appears before \textit{<d>} to form the correct grapheme sequence \textit{<c><o><n><s><t><r><u><c><t><e><d>}. Rather than presenting this as a static correction, the system visually highlights these changes to draw Matilda’s attention to how sound-letter patterns map across morphology and phonology.

Finally, \system prompts Matilda to suggest words that share the same base \textit{<struct>} (Fig.~\ref{fig:ux}C1). She types \textit{structure} and \textit{insstruct} (Fig.~\ref{fig:ux}C2). The system affirms \textit{structure} but flags \textit{insstruct} for having an extra \textit{<s>}, explaining the morphological constraint (Fig.~\ref{fig:ux}C3). This final step reinforces her understanding of morphological families and related word forms before she returns to her essay.

Through these brief, multimodal interactions--speaking, dragging, typing---Matilda not only corrects her mistake but also deepens her understanding of word structure and related forms. By condensing inquiry into a short, bounded sequence, the system allows her to engage in a "micro-lesson" without significantly derailing her drafting process. This interaction model illustrates how \system seeks to balance writing fluency with learning, turning selected spelling errors into manageable opportunities for structured linguistic exploration.

%% file: 04_DSL.tex
\section{Pedagogical Program Language}
\label{sec:DSL}
To develop \system, we need to encode the dynamic conversational practice of SLPs into something a computer can assemble in real time from a single misspelling and its context. Our pedagogical programming language (i.e., DSL) formalizes how SLPs reason, question, and guide discovery during therapy as composable program primitives that can be synthesized into concise, adaptive instructional sequences. Crucially, our DSL is grounded in practitioner concepts and built with practitioners. We first inventoried canonical SWI moves and terminology from established texts (books written by expert practitioners, instructional material used in graduate programs, and skill development workshops~\cite{hegland2021beneath,barnett2024vocabulary,bowers2006teaching,constantine2024structured,loveless2024high}) into a spreadsheet. Next, working with a practicing SLP, who is also a co-author, we organized the moves as program primitives, including language primitives, hypothesis templates, action primitives, and control and compositional primitives. For this collaborative work, we utilized digital whiteboards to organize the concepts, created sketches of program logic and instructional moves, and also produced storyboards of envisioned learning episodes. This was an iterative process that involved weekly virtual meetings with the research team, spanning over six months. During this process, we also consulted with two other practicing SLPs who have over 10 years of experience to receive feedback on the evolving DSL. 

In what follows, we describe each category of primitives that form the basis of our DSL. Collectively, these primitives keep the logic and vocabulary of practice intact while making them computationally operable for just-in-time, in-context human learning.

\subsection{Language Knowledge Primitives}
This set of primitives defines the linguistic substrate that each pedagogical program operates on. Concretely, they represent the structured linguistic information an SLP implicitly draws upon when reasoning about a misspelling, including knowledge of a word's morphology, graphemes, phonemes, word families, and etymology. In our DSL, this knowledge is encoded as a structured record \texttt{word\_properties} whose named attributes (shown in Table~\ref{tab:language_primitives}) correspond directly to the analytic categories used in SWI. Just as with SLPs, these attributes enable the pedagogical programs to reason about the logic behind a spelling and pose the same kinds of questions a practitioner might ask, such as "What is the base of the word?", "What sound do these letters represent?", and "Which other words share this base?"

\begin{table*}[hbt]
    \centering
    \small
    \renewcommand{\arraystretch}{1.5}
    \arrayrulecolor{black}
    \setlength{\fboxsep}{0pt}
    \setlength{\fboxrule}{0.3pt}
    \fbox{%
    \begin{tabular}{@{\hspace{5pt}}l|>{\raggedright\arraybackslash}p{6cm}|>{\raggedright\arraybackslash}p{5cm}@{\hspace{5pt}}}
        \textbf{DSL Attribute} & \textbf{Pedagogical Role in SLP Instruction} & \textbf{Example (``reaching'')} \\
        \hline
        \tagbox{S2Color}{\texttt{morphemes}} & Meaning-bearing parts composing the word; focus of ``box the base'' and word-sum construction. & \texttt{[``reach'', ``-ing'']} \\
        \hline
        \tagbox{S2Color}{\texttt{bases}} & Core lexical root carrying meaning across a family; anchors family-based pattern recognition. & \texttt{["reach"]} \\
        \hline
        \tagbox{S2Color}{\texttt{prefixes}} & Bound morphemes before the base; invite prefix-meaning inquiries and family extension. & \texttt{none} \\
        \hline
        \tagbox{S2Color}{\texttt{suffixes}} & Bound morphemes following the base; used in suffixing-convention and inflectional change prompts. & \texttt{[``-ing"]} \\
        \hline
        \tagbox{S2Color}{\texttt{graphemes}} & Letter or letter-team units representing phonemes; supports grapheme identification and comparison. & \texttt{[``r'', ``ea'', ``ch'', ``i'', ``ng'']} \\
        \hline
        \tagbox{S2Color}{\texttt{phonemes}} & Sound units of the spoken form; validate or contrast the learner's sound--spelling reasoning. & \texttt{[\textipa{``/r/'', ``/i:'', ``/tS/'', ``/I/'', ``/N/''}]}  \\
        \hline
        \tagbox{S2Color}{\texttt{related\_words}} & Morphological or orthographic relatives for matrices and ``in/out'' sorting; surface consistent base spelling. & [ [ ``reach'', ``reaching'', ``overreach'', [``teach'', ``preach'' ] ]\\
        \hline
        \tagbox{S2Color}{\texttt{etymology}} & Historical origin explaining modern spelling stability; supports etymological reasoning. & ``Old English \emph{reccan} (`to stretch out')'' \\
        \hline
        \tagbox{S2Color}{\texttt{homophones}} & Same sound, different spelling or meaning; prompts meaning-based distinction. & \texttt{none} \\
        \hline
        \tagbox{S2Color}{\texttt{semantic\_appropriateness}} & Ensures the written word matches intended meaning; drives meaning-first inquiry when mismatched. & ``Intended `extending or stretching out'(She is reaching for the shelf)'' \\
        \hline
        \tagbox{S2Color}{\texttt{syntactic\_correctness}} & Checks grammatical fit; distinguishes morphological from typographic errors. & ``Progressive verb fits: `She is reaching\ldots' '' \\
        \hline
        \tagbox{S2Color}{\texttt{meaning\_understood}} & If unclear, prioritizes definition/denotation prompts before structure/GPC work. & ``Learned typed `reeching' intended `reaching\ldots' '' \\
        \hline
        \tagbox{S2Color}{\texttt{context\_sentence}} & Provides authentic sentence evidence; used in meaning or boundary reasoning. & ``She was reeching for the book.'' \\
    \end{tabular}%
    }
    \vspace{3pt}
    \caption{List of Language knowledge primitives encoded in \texttt{word\_properties}.}
    \label{tab:language_primitives}
\end{table*}

\subsection{Hypotheses Template Primitives}
Our DSL formalizes the inquiry logic of structured word inquiry as a set of reusable \textbf{hypothesis templates}. Each template represents a recurring pattern of \textit{reasoning} that enables the learner to test and refine a hypothesis about how a word works. It expresses a single pedagogical move typical in SWI as a structured \textit{quintuple} consisting of preconditions, evidential features, action base, warrant, and learning effect. \textit{Preconditions} state when a line of inquiry is warranted. They are conditional predicates over an observable situation, including the type of deviation or error in a learner's spelling production, the linguistic characteristics of the intended word, and basic usage context, such as semantic appropriateness. Concretely, they are represented as a combination of Boolean predicates computed over the linguistic primitives as well as diagnostic descriptors in natural language. The predicate provides a formal check for whether the template applies, while the descriptor captures practitioner judgment that is hard to encode as a formal rule. 

For instance, in the grapheme identification template (H8 in Table~\ref{tab:hypothesis_templates_abbrev}), the symbolic guard checks that the that the deviation is \textit{grapheme–local} rather than structural: morphology and boundary errors are not implicated (\texttt{$\neg$(prefix\_error $\vee$ suffix\_error $\vee$ segmentation\_error)}), no suffixing convention applies (\texttt{$\neg$ suffixing\_change\_applies}), the phoneme sequence of the attempt essentially matches the target (\texttt{phoneme\_match} $\le \varepsilon$), the mismatch is small and local (\texttt{grapheme\_mismatch\_count} $\in \{1,2\}$), and morpheme boundaries are preserved (\path|morpheme_boundaries_preserved|). The paired diagnostic descriptor states this in practitioner terms e.g., ``sounds correct but uses a nonstandard letter team.'' Thus \emph{reech}→\emph{reach} satisfies the guard (phonology intact, one letter–team substitution), whereas \emph{runing}→\emph{running} is excluded by the suffixing-change check and \emph{alot}→\emph{a lot} by the segmentation predicate. At runtime, an LLM assigns soft confidence to descriptors, allowing preconditions to serve as interpretable constraints and probabilistic cues for selecting the most pedagogically relevant hypothesis.

\textit{Evidential features} specify what linguistic features must be inspected in order to carry out the inquiry. In our DSL, evidential features are typed references to the linguistic primitives listed in Table~\ref{tab:language_primitives}. For instance, for grapheme inquiry, the evidence is the pair of grapheme inventories and their alignment to the phoneme sequence; Crucially, evidential features do \textit{not} enumerate examples themselves; they expose the handles that the \textit{program synthesis} stage will later use to instantiate a small contrast set. For grapheme inquiry, for instance, the evidence is just the learner and target grapheme sequences aligned to the same phoneme sequence; during synthesis, these handles permit selecting minimal contrasts for the contested mapping (e.g., attested spellings of /i/ such as <ee>, <ea>). Analogously, for suffixing, evidence is the base and the suffix with relevant orthographic context, etc. In short, evidential features say ``what to look at'' and program synthesis later decides which concrete examples to show.

The \textit{Action Base} describes the type of learner moves, i.e., established spelling interventions to carry out the inquiry. Typical action types include identifying graphemes in one's own spelling, mapping phonemes to letter teams, boxing the base, analytically removing a suffix, constructing a forward word-sum, assembling a compact family matrix, and sorting IN/OUT relatives, or orally segmenting a run-together token. Different interfaces might realize the action types differently (e.g., highlight or tap to identify graphemes, drag-to-align or speak-to-align for phoneme$\rightarrow$grapheme mapping) while preserving identical pedagogical semantics. The \textit{Learning Effect} records the conceptual outcome that the move is designed to produce, i.e., an update to the learner's knowledge state. Typical effects include meaning clarified, base verified, word-sum assembled or deconstructed, convention induced, boundaries corrected, grapheme–phoneme mapping aligned, family confirmed, etymological rationale established, homophone distinguished, or visual discrimination achieved. These intended learning effects not only describe the outcome of the current inquiry but also serve as preconditions for subsequent hypotheses and as features for downstream selection without committing to any particular control policy. 

Finally, \textit{Warrants} name the explanatory link that connects the action's observables to the learning effects. Formally, it is an explanatory operator with a type (e.g., grapheme comparison) and parameters drawn from the evidential features (which items to contrast, which rule to name, which relatives to surface). Crucially, the warrant is what generates the \emph{just-in-time conversation} for guided inquiry toward the learning effect. Because the warrant specifies the explanatory move and the slots it fills, the system can produce the next utterance that validates the learner's work, introduces the right contrast, and generalizes the insight. For \textit{grapheme identification} (H8), once the learner has identified letter teams, the warrant instantiates a comparison: ``Now line up your teams with the target; notice that \textit{<ea>} is the stable team in this family—so here we write \textit{reach}.'' When the observed evidence does not support the intended explanation, the warrant also determines the conversational pivot (e.g., from comparison to meaning-first or to relatives), keeping the guidance evidential rather than corrective.

\subsection{Compositional Primitives}
This set of primitives describes how discrete acts of inquiry unfold coherently over time. In structured word inquiry, learning progresses flexibly, where the learner advances a hypothesis, inspects evidence, articulates what that evidence reveals, and then extends or closes the inquiry. In \system, our goal is that given the current misspelling (i.e., attempt state) and the intended canonical form (i.e., target state), a plan is generated in the form of a sequence of inquiry steps, in which each step yields a learning effect that reduces some part of the observed discrepancy or produce the specific evidence needed for the next justified step. In effect, the plan serves as a small \textit{proof-by-inquiry} that bridges the gap from attempt to target. 
\textit{Compositional} primitives specify the legality of moves as dependencies among the hypothesis template fields. In a simple \emph{sequence}, meaning clarification can precede base identification; grapheme identification naturally precedes a grapheme comparison; verification of a base licenses testing a suffixing condition. \emph{Nesting} reflects the way SLPs zoom in: a suffixing test sits inside a word-sum construction; verifying likely relatives sits inside a family-pattern generalization. \emph{Branching} captures the exploration of alternatives SLPs routinely entertain: try two plausible vowel teams or contrast two related families, then merge back once one branch yields a learning effect or both are ruled out. \emph{Closure} occurs when the discrepancies relevant to the learner's attempt have been resolved or when the inquiry has produced the named understanding that the SLP was aiming for. 

Collectively, these primitives let \system treat instruction as constrained search under uncertainty. Given an attempt state and a target state, the synthesizer explores the legal space defined by the control and compositional primitives. It proposes short sequences of hypothesis templates whose preconditions are satisfied, whose actions bind the needed evidence, whose warrants generate just-in-time explanations, and whose learning effects shrink the discrepancy. The following section presents our program-synthesis architecture that operationalizes this process.

\begin{table*}[hbt]
    \centering
    \renewcommand{\arraystretch}{1.4}
    \arrayrulecolor{black}
    \setlength{\fboxsep}{0pt}
    \setlength{\fboxrule}{0.3pt}
    \fbox{%
    \begin{tabular}{@{\hspace{4pt}}c|>{\raggedright\arraybackslash}p{3.3cm}|>{\raggedright\arraybackslash}p{3.6cm}|>{\raggedright\arraybackslash}p{3.65cm}|>{\raggedright\arraybackslash}p{3.65cm}@{\hspace{4pt}}}
        \textbf{ID} & \textbf{Template} & \textbf{Key Preconditions} & \textbf{Action Base} & \textbf{Learning Effect} \\
        \hline
        \tagbox{S2Color}{H1} & Meaning Clarification & Meaning unclear or semantically inappropriate. & Define intended meaning; contextual prompt. & Meaning aligned; enables next inquiry. \\
        \hline
        \tagbox{S2Color}{H2} & Base Denotation & Base and affix present; meaning known. & Identify main meaning part. & Semantic core anchored in base. \\
        \hline
        \tagbox{S2Color}{H3} & Morphological Structure & 1 or more affixes; single base. & ``Box the base'' or decompose parts. & Base verified; structure understood. \\
        \hline
        \tagbox{S2Color}{H4} & Word-Sum Construction & Decomposable into morphemes. & Build word step-by-step. & Morphemes combined $\rightarrow$ correct form. \\
        \hline
        \tagbox{S2Color}{H5} & Spelling Convention & Suffixing rule applies (e.g., doubling, e-drop). & Inspect suffix rule via word-sum. & Orthographic rule induced. \\
        \hline
        \tagbox{S2Color}{H6} & Word Matrix & Base stable across 2 or more relatives. & Assemble word matrix grid. & Family pattern generalized. \\
        \hline
        \tagbox{S2Color}{H7} & Word Boundaries & Segmentation or spacing error. & Segment aloud, retype spaced form. & Correct word boundaries restored. \\
        \hline
        \tagbox{S2Color}{H8} & Grapheme Identification & 2 or less grapheme mismatch. & Identify graphemes, compare to target. & Grapheme--phoneme alignment achieved. \\
        \hline
        \tagbox{S2Color}{H9} & Etymology Inquiry & Historical root available. & Trace origin; discuss spelling stability. & Origin-based spelling rationale. \\
        \hline
        \tagbox{S2Color}{H10} & Morphological Relatives & Base shared by other words. & Sort IN/OUT relatives; construct family. & Base spelling reinforced by family. \\
        \hline
        \tagbox{S2Color}{H11} & Morpheme Verification & Base/affix uncertainty. & Verify morphemes across known words. & Morpheme consistency confirmed. \\
        \hline
        \tagbox{S2Color}{H12} & Etymological Relatives & Common ancestor offers insight. & Compare historical cousins. & Etymological family explained. \\
        \hline
        \tagbox{S2Color}{H13} & False Relatives & Only visual similarity. & Contrast look-alike vs.\ true relatives. & False relatives excluded. \\
        \hline
        \tagbox{S2Color}{H14} & Phoneme--Grapheme Map & Phonology is similar to target. & Map sounds to graphemes. & Sound--spelling map aligned. \\
        \hline
        \tagbox{S2Color}{H15} & Pronunciation Variation & Family shows sound shifts. & Compare relatives with shared grapheme. & Stable spelling despite sound change. \\
        \hline
        \tagbox{S2Color}{H16} & GPC Pattern Consistency & Consistent spelling, varied pronunciation. & Compare family spellings. & Pattern consistency generalized. \\
        \hline
        \tagbox{S2Color}{H17} & Homophone Distinction & 1 or more homophones exist. & Sort by meaning; contrast forms. & Correct form tied to meaning. \\
        \hline
        \tagbox{S2Color}{H18} & Similar Word Comparison & Similar orthography detected. & Visual contrast task. & Subtle form difference noticed. \\
    \end{tabular}%
    }
    \vspace{3pt}
    \caption{Summarized Hypothesis Templates of \system.}
    \label{tab:hypothesis_templates_abbrev}
\end{table*}

%% file: 05_system.tex
\section{System Architecture}


\begin{figure}
    \centering
    \includegraphics[width=0.8\linewidth]{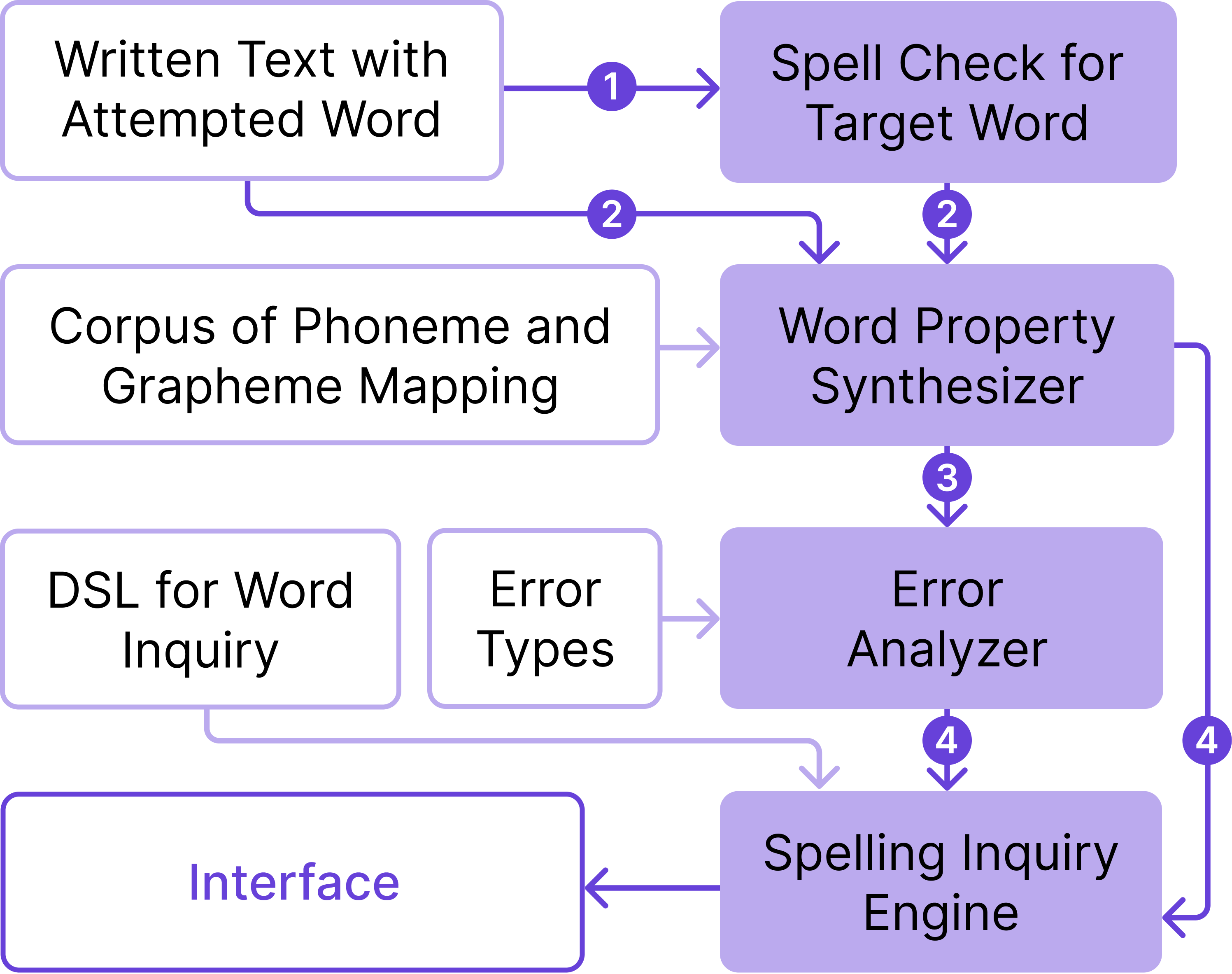}
    \caption{Overview of the System Architecture of \system. The number labels on the edges mark the timestep at which each module is activated.}
    \label{fig:sys_arc}
\end{figure}

\begin{figure*}
    \centering
    \includegraphics[width=1\textwidth]{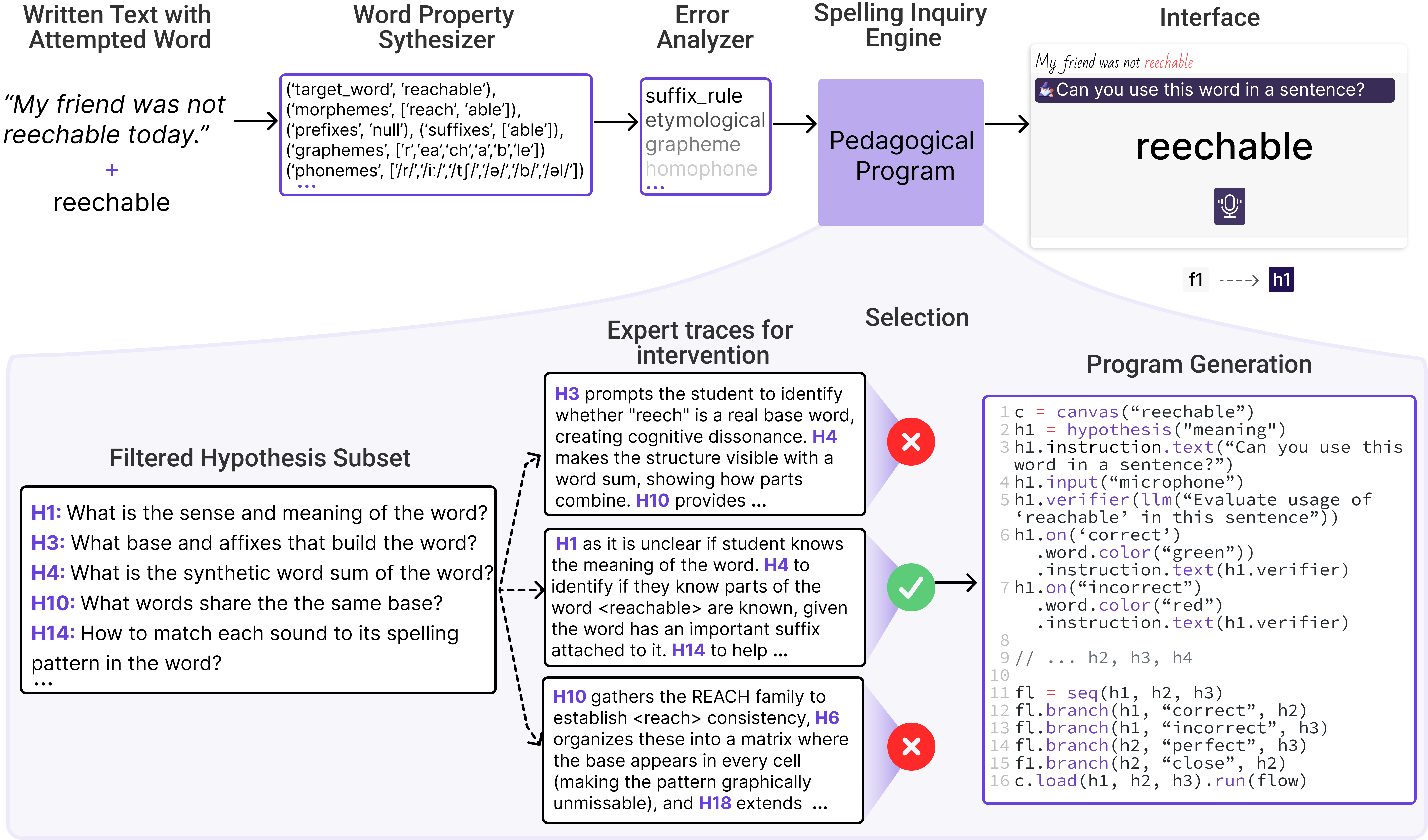}
    \caption{This figure showcases an example for the pipeline, providing each step's output.}
    \label{fig:sys_arc_example}
\end{figure*}

\begin{figure*}
    \centering
    \includegraphics[width=0.9\textwidth]{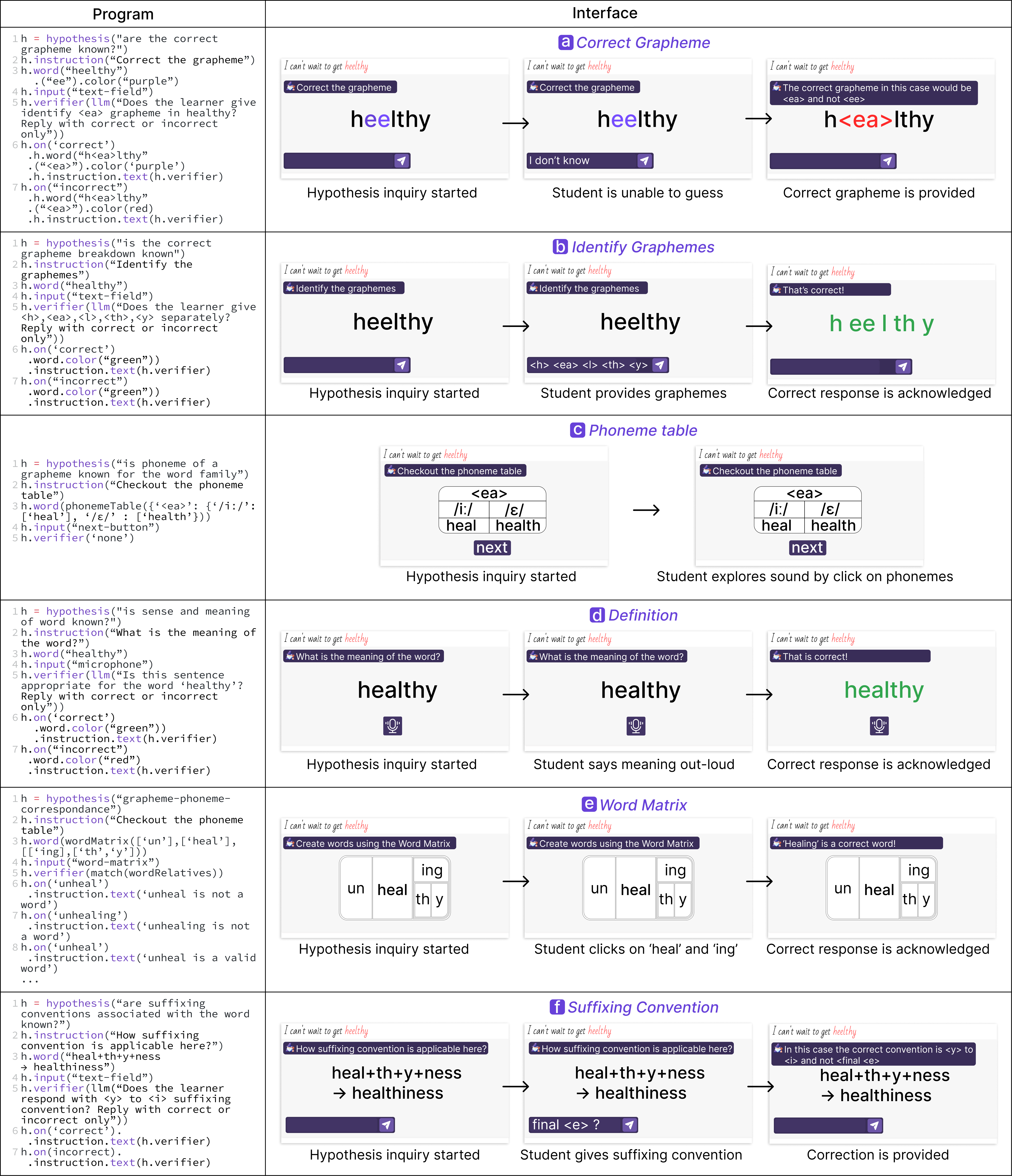}
    \caption{This figure showcases some example hypothesis testing in the interface (right) and the program that guides that interface (left).}
    \label{fig:examples}
\end{figure*}

The pedagogical program language presented in Section 4 defines what constitutes valid SWI instruction—the hypotheses, evidence, and warrants that structure expert practice. This section describes how these abstract specifications are operationalized at runtime. The system architecture translates the DSL's declarative primitives into executable instructional sequences, automatically synthesizing programs from observed spelling errors and rendering them as interactive interventions within the writing interface.

This architecture addresses a fundamental limitation in existing spelling tools. Conventional spell checkers detect errors and suggest corrections based on string similarity~\cite{kukich1992techniques}, treating spelling as a mechanical transcription problem. \system must instead infer why a misspelling occurred, what it reveals about the learner's linguistic reasoning, and which line of inquiry will advance their orthographic understanding. The spelling inquiry engine, implemented in Python, orchestrates this process through five modules: a spell checker identifies target words from context (Fig.~\ref{fig:sys_arc}, step 1), a word property synthesizer extracts morphological and phonological features (Fig.~\ref{fig:sys_arc}, step 2), an error analyzer diagnoses likely causes (Fig.~\ref{fig:sys_arc}, step 3), a spelling inquiry engine composes hypothesis templates into executable programs following the DSL notation (Fig.~\ref{fig:sys_arc}, step 4), and an interface generator renders these programs as interactive components. Together, these modules transform the DSL's declarative specifications into adaptive, just-in-time instruction tailored to individual learner errors.

\subsection{Spell Checker}
The spell checker is the entry point of \system, responsible for detecting misspellings and identifying their most likely target words. It operates continuously in the background and is activated after every second long pause in writing. Upon activation, it analyzes the surrounding sentence to locate potential spelling errors and predict their intended forms. For example, when a learner writes “Reechable,” the spell checker identifies the error and generates “Reachable” as the target word. We designed this module to address the limitations of conventional spell checkers, which rely primarily on string similarity and often fail when a learner’s attempt deviates substantially from the correct form (e.g., yous → use). Even commercial tools like Grammarly~\cite{grammarly2025} tend to substitute alternative word choices rather than identifying the learner’s intended word. Because \system aims to reason about the learner’s linguistic understanding, it requires target predictions that are both contextually grounded and semantically faithful. To achieve this, the spell checker uses a large language model (LLM) to perform contextual error detection and correction. The LLM considers both the local sentence and the broader discourse context to infer the intended word rather than relying solely on edit distance. Once the target word is identified, both the attempted and target forms—along with their sentence context—are passed to the next module, the word property synthesizer, which begins the process of linguistic analysis.

\subsection{Word Property Synthesizer}
The word property synthesizer extracts the linguistic features of the target word that are relevant for analysis and instruction. Specifically, it identifies morphological, phonological, and orthographic properties such as morphemes (meaning-bearing units), graphemes (letters or letter teams representing sounds), and phonemes (sound units). These features serve as the linguistic foundation on which the system builds hypotheses about the learner’s spelling behavior. We designed this module to capture the kinds of linguistic reasoning speech-language pathologists (SLPs) use when teaching spelling. Standard spell checkers only produce surface-level corrections, but effective instruction requires understanding why a word is spelled the way it is—its structure, sound patterns, and etymology. To operationalize this expert reasoning, the synthesizer must produce interpretable, structured representations of words that downstream modules can reason over. To achieve this, the word property synthesizer uses a large language model (LLM) prompted with a structured specification and few-shot examples. It receives the target word and its writing context, then outputs a set of linguistic predicates in the form of \textit{(field-name, value)} tuples (e.g., \textit{('morpheme', 'reach')}, \textit{('suffix', 'able')})—instantiating the language knowledge primitives defined in Table 1. A corpus of English phoneme–grapheme mappings is provided as grounding context to ensure validity. The resulting property set—representing the linguistic knowledge an SLP would reference—is then passed to the error analyzer for diagnosing the underlying causes of the misspelling.

\subsection{Error Analyzer}
The error analyzer identifies the most plausible causes of a learner’s misspelling. It takes as input the linguistic properties generated by the word property synthesizer and compares them against an annotated list of common error types, such as phoneme–grapheme mismatches, morphological confusions, or suffixing errors. This module bridges linguistic analysis and instructional reasoning by explaining why the error occurred rather than merely labeling it. Understanding the underlying cause is essential for selecting pedagogically meaningful interventions. To achieve this, the error analyzer uses a large language model (LLM) prompted with the annotated error taxonomy and few-shot examples. The LLM evaluates how the learner’s attempted and target words diverge across linguistic dimensions and returns a ranked list of likely error categories. This structured output is then passed to the spelling inquiry engine for generating the corresponding guided intervention program.

\subsection{Spelling Inquiry Engine}
The spelling inquiry engine instantiates the pedagogical program language described in Section 4, generating concrete intervention programs by composing hypothesis templates into executable sequences that \system delivers to the learner. It determines what instructional steps to present and in what order, transforming linguistic analyses and diagnosed error types into a coherent sequence of inquiry activities. Each program represents a short cycle of hypothesis testing designed to help the learner reason about the structure, sound, or meaning of the target word. We designed this module to model how speech-language pathologists (SLPs) guide learners through discovery-based spelling instruction. Rather than offering fixed feedback, the engine composes adaptive instructional sequences that reflect pedagogical intent—selecting, sequencing, and framing interventions based on the learner’s observed error and linguistic context. To achieve this, the engine integrates symbolic reasoning with language-model inference. Implemented in Python, it converts the filtered set of hypotheses into a structured program using our domain-specific language (DSL), which defines the syntax and logic of instructional moves. The resulting program is serialized and rendered in React as an interactive interface. As shown in Figure~\ref{fig:sys_arc_example}, the engine consists of four submodules: (A) Hypothesis Filter, (B) Trace Generator, (C) Trace Selector, and (D) Program Generator, which together simulate the expert reasoning process of an SLP.

\subsubsection{Hypothesis Filter}
\label{sec:filter}
The hypothesis filter performs the first selection step in the spelling inquiry engine. It narrows the full set of pedagogical hypotheses to those that are relevant for the learner’s specific word and error context. Each hypothesis encodes a possible instructional focus (Section 4.2)—such as analyzing a base word, exploring a grapheme–phoneme mapping, or inspecting a suffixing convention. This filtering process ensures that the system considers only meaningful and non-trivial inquiries. Without it, the program could surface redundant or pedagogically irrelevant interventions (e.g., asking about suffixing when none applies). The goal is to approximate how an SLP intuitively rules out lines of inquiry that do not advance the learner’s understanding. To achieve this, the filter uses a unification-based pattern matcher to align the predicate tuples from the word property synthesizer with the preconditions defined in each hypothesis. Guards and diagnostic descriptors encoded in the hypotheses are evaluated symbolically or semantically using a language model. The output is a refined set of hypotheses that satisfy all constraints, forming the candidate pool for trace generation in the next stage.

\subsubsection{Trace Generator \& Trace Selector}
The trace generator and trace selector together form the core reasoning mechanism of the spelling inquiry engine. Their role is to construct and refine instructional sequences—called traces—that guide the learner through 2–5 hypothesis-driven steps of exploration. The generator proposes multiple alternative instructional paths, while the selector evaluates and ranks these paths to identify the one most pedagogically sound. This two-phase design is modeled after human inductive reasoning and draws inspiration from the hypothesis search framework~\cite{wang2023hypothesis}, which separates divergent generation from convergent evaluation. Effective instruction requires both creativity and precision: educators first imagine several possible ways to guide a learner, then choose the one that best advances understanding. By mirroring this cognitive process, \system balances exploration (diversity of ideas) with instructional rigor (quality of reasoning).

To achieve this, we implement a multi-instance LLM architecture. The Trace Generator launches three parallel LLM instances, each independently exploring distinct instructional traces grounded in the learner’s linguistic and error profiles. This parallelism promotes divergent reasoning, surfacing multiple pedagogically plausible sequences that link the attempted and target words through morphological or phonological insight. The resulting traces are then passed to the Trace Selector, a single LLM instance that performs convergent reasoning. It reviews and compares the candidate traces using criteria such as pedagogical validity, conceptual coherence, and instructional clarity. The top-ranked trace—the one most likely to foster meaningful understanding—is selected and forwarded to the Program Synthesizer for conversion into executable DSL code. This architecture ensures both breadth and depth: breadth through generative exploration of multiple possible instructional paths, and depth through rigorous evaluation and selection. Together, these modules enable \system to emulate the reflective judgment of an expert SLP—creatively hypothesizing instructional strategies and critically choosing the most effective one to guide the learner’s inquiry.

\subsubsection{Program Synthesizer}
The program synthesizer converts the selected instructional trace into an executable pedagogical program following the DSL notation defined in Section 4. It operationalizes the reasoning captured in hypothesis templates—their preconditions, evidential features, actions, warrants, and learning effects—transforming abstract hypotheses into concrete, interactive steps that guide the learner through linguistic inquiry. Each synthesized program specifies instructional actions, learner inputs, and conditional feedback, following the notation defined in our domain-specific language (DSL) (see “Generated Program” in Figure~\ref{fig:sys_arc_example}).

We designed this module to bridge pedagogical intent and interface behavior. While the preceding modules determine what to teach and why, the synthesizer determines how that reasoning manifests interactively. It ensures that every instructional step—such as asking the learner to identify a base or compare graphemes—is both executable and adaptable to learner responses. Technically, the synthesizer accepts the best trace, word properties, and DSL schema as input context. Using structured prompting, it generates a declarative program with branching logic that adapts dynamically to user input. Each branch encodes conditional feedback and next-step transitions, enabling the interface to respond to learner actions in real time. To ensure reliability, the synthesizer validates the generated program for syntax and runtime consistency; if compilation fails, the system automatically regenerates it. The resulting program is serialized and passed to the interface generator for rendering.

\subsection{Interface Generator}
The interface generator renders the synthesized pedagogical program as an interactive learning experience. It translates the declarative program produced by the synthesizer into dynamic interface components that guide the learner through inquiry-based spelling correction. Each hypothesis node in the program corresponds to a visual interaction module that binds instructional text, learner input, and feedback logic, allowing the system’s reasoning to unfold directly within the writing environment.

We designed this module to preserve an isomorphic relationship between pedagogical logic and interface behavior. Rather than displaying static feedback, the interface visually manifests the structure of the instructional program—showing learners not just the correction, but the reasoning process behind it. This design enables the same program to operate seamlessly in both headless (text-based) and interactive (graphical) modes, making it adaptable across different learning contexts. Technically, the interface generator parses the JSON execution plan produced by the program synthesizer and compiles it into React components. Each component corresponds to a hypothesis in the instructional trace (e.g., \texttt{h1}, \texttt{h2}, \texttt{h3}), specifying its instructional text, input affordances (such as text fields or visual tools like paint-rollers), and verification conditions. Conditional branches (\texttt{on(true)}, \texttt{on(false)}) define how the interface adapts to learner responses in real time, ensuring pedagogical continuity across different interaction outcomes. This direct coupling between program logic and user interaction enables \system to render the reasoning process of an SLP as an interactive learning experience.

\subsection{Implementation Details}
\system integrates a Python backend with a React frontend. The spelling inquiry engine is implemented in Python using a unification-based pattern matcher~\cite{maclellan_pyplan} and Anthropic’s Claude Sonnet~\cite{anthropic_claude_sonnet45} for hypothesis filtering, generation, and selection. From text analysis through intervention generation, the pipeline makes 8 LLM API calls total, including parallel calls for hypothesis search and final DSL synthesis; using Claude Sonnet 4.5, this costs approximately \$0.70 per complete word inquiry. The complete pipeline—from text analysis to intervention generation—involves 8 LLM API calls, including parallel threads for hypothesis search. During testing with Sonnet 4.5, this averaged \$0.70 per complete word inquiry (excluding optional speech services). The synthesized instructional programs are serialized as JSON execution plans, which the React interface renders as interactive components mirroring the system’s pedagogical logic. To support multimodality, \system employs Google Gemini’s speech-to-text for capturing spoken responses and Gemini’s text-to-speech for verbalizing instructions and feedback, enabling voice-based interaction alongside text input. This modular setup ensures consistent execution across conversational and interactive modes while remaining adaptable for integration with future writing or tutoring systems.

%% file: 06_tech_evaluation.tex
\section{Technical Evaluation}
To validate our technical pipeline, we conducted an expert evaluation study with specialists in structured word inquiry and spelling instruction for children.

\subsection{Material}
We collected 20 writing samples from children in grades 2-5, each containing one or more spelling errors. Using the \system pipeline, we generated 50 conversations between a simulated expert tutor and simulated student, each addressing one misspelled word with 2-5 just-in-time interventions. To ensure the evaluation workload remained manageable for the experts, the 50 conversations were randomly divided into two sets of 25 distinct transcripts.
For each intervention, the system provided: (1) the targeted SWI question type, (2) the selected hypothesis, (3) the rationale, and (4) the instructional dialogue. The system's responses dynamically adjusted to simulated learner responses following the branching logic expressed in our domain-specific language.

\subsection{Participants}

We recruited 10 experts through word-of-mouth and mailing lists. The demographic information pertinent to each expert are present in the table~\ref{tab:participants}. The number of participants is in line with prior work that has required domain- experts/teachers~\cite{gong2025callisense,salminen2024deus}.

\subsection{Study Procedure}
Experts first attended a Zoom orientation session explaining the evaluation task. They then rated conversation transcripts through a custom web interface using 5-point Likert scales for each intervention: (Q1) ``The action selected is appropriate''; (Q2) ``The reason for action selection is valid''; (Q3) ``I feel confident in my assessment.'' Experts could also add qualitative notes. After completing a guided transcript, participants independently rated 25 transcripts from their randomly assigned set over multiple days. Completing the task required 2-6 hours. Participants received \$50 compensation.

\subsection{Results}
Experts responded to 109 interventions in the first set and 105 in the second set. We report results jointly for both sets that were given to 5 participants each. This is because the words present between the two sets were of similar difficulty. Additionally, there were no statistical difference across the sets. 

\subsubsection{Pedagogical Reasoning Validity (Q1).} 
Experts demonstrated exceptionally high consensus on the soundness of the system's pedagogical reasoning, with 74.7\% perfect agreement across all ten raters ($\mu=4.90$, $\sigma=0.47$). The near-ceiling ratings indicate strong expert agreement that the system consistently provided appropriate pedagogical explanation for student errors.

\subsubsection{Instructional Action Appropriateness (Q2).} 
Actions received consistently positive rating ($\mu=4.32$, $\sigma=1.13$) with moderate inter-rater agreement of 26.7\% perfect agreement. The higher variance reflects the pedagogically complex nature of tutoring, where multiple valid instructional approaches often exist for addressing student errors. Qualitative notes revealed that disagreement primarily centered on three topics: (1) 1) pedagogical sequencing (e.g., whether to address morphology before grapheme-phoneme correspondence), (2) depth of explanation appropriate for different age groups, and (3) terminology preferences (e.g., ``base'' vs. ``base word,'' ``vowel digraph'' vs. ``vowel team'').

\subsubsection{Expert Confidence (Q3).} Raters reported high confidence in their evaluations ($\mu=4.92$, $\sigma=0.47$), along with high agreement of 80.1\%, validating the reliability of the expert assessment process.

\subsubsection{Interpretation.}
The contrasting patterns of expert agreement reveal two important characteristics of the system. The high agreement on reasoning validity (74.7\% perfect agreement) in Table~\ref{tab:rater-agreement}) demonstrates that the system exhibits strong pedagogical competence---it consistently generates sound rationales that experts recognize as appropriate. In contrast, the moderate agreement on instructional actions reflects genuine pedagogical flexibility rather than a system limitation. Effective tutoring inherently admits multiple valid approaches, and expert teachers frequently disagree on optimal instructional sequences even while concurring on underlying pedagogical principles.Experts also noted that the simulated dialogue mirrored their own teaching strategies, praising the flexibility and breadth of generated interventions. Raters specifically commended the system's orthographic precision, such as correctly identifying the <-ion> suffix in ``action'' rather than the common error <-tion>.

\begin{table}[hbt]
    \centering
    \small
    \renewcommand{\arraystretch}{1.5}
    \arrayrulecolor{black}
    \setlength{\fboxsep}{0pt}
    \setlength{\fboxrule}{0.3pt}
    \fbox{%
    \begin{tabular}{@{\hspace{5pt}}c|c|c|c@{\hspace{5pt}}}

        \textbf{ID} & \textbf{Age} & \textbf{Spelling Instruction} & \textbf{SWI Experience} \\
         & & \textbf{Experience} & \\
        \hline
        E1 &  71 & 20+ years & 8 years \\
        \hline
        E2 &  34 & 9 years & 2 years \\
        \hline
        E3 &  55 & 15 years & 4 years \\
        \hline
        E4 &  47 & 15 years & 8 years \\
        \hline
        E5 &  47 & 20+ years & 7 years \\
        \hline
        E6 &  56 & 14 years & 7 years \\
        \hline
        E7 &  51 & 20+ years & 10 years \\
        \hline
        E8 &  34 & 9 years & 2 years \\
        \hline
        E9 &  56 & 13 years & 5 years \\
        \hline
        E10 &  60 & 15 years & 13 years \\
    \end{tabular}%
    } 
    
    \vspace{5pt}

    \caption{Participant Demographics and Experience}
    \label{tab:participants}
\end{table}

\begin{table*}[hbt]
    \centering
    \renewcommand{\arraystretch}{1.5}
    \arrayrulecolor{black}
    \setlength{\tabcolsep}{8pt}
    \setlength{\fboxsep}{0pt}
    \setlength{\fboxrule}{0.3pt}
    \fbox{%
    \begin{tabular}{@{\hspace{5pt}}l|c|c|c|c|c@{\hspace{5pt}}}
        \textbf{Dimension} & \textbf{Mean (SD)} & \textbf{\% $\ge$ 4} & \textbf{\% = 5} & \textbf{t-test vs Neutral ($\mu$=3)} & \textbf{Perfect Agreement} \\
        \hline
        Reasoning & 4.90 (0.47) & 97.2\% & 94.4\% & 41.97* & 74.7\% \\
        \hline
        Action & 4.32 (1.13) & 82.0\% & 70.2\% & 9.59* & 26.7\% \\
        \hline
        Confidence & 4.92 (0.47) & 97.5\% & 95.7\% & 43.04* & 80.1\% \\
    \end{tabular}%
    }
    
    \raggedleft
    \footnotesize \vspace{4pt} * p < 0.001
    \vspace{3pt}
    \caption{Inter-rater agreement and rating distributions across evaluation dimensions. \% $\geq$4 and \% =5 indicate proportions rating "agree" or higher and "strongly agree," respectively. Participant-level t-tests compare means against the neutral midpoint ($\mu$=3).}
    \label{tab:rater-agreement}
\end{table*}

%% file: 07_user_evaluation.tex
\section{Preliminary Feedback from Children}
We present findings from our preliminary usability evaluation with 7 children, focusing on their experience with the spelling tool and overall usability perceptions.

\subsection{Participants}
We recruited 7 children (ages 7--11, grade 2--5) through word-of-mouth and mailing lists. Participant demographics included 6 native English speakers and 1 native Russian speaker, all currently learning to spell in English. When asked about their familiarity with structured word inquiry, no participant reported prior exposure to this approach. Parents reported varied spelling difficulty levels for their children (range 1--5 on a 5-point scale, median = 3), indicating a diverse sample in terms of spelling proficiency.

\subsection{Study Procedure}
We conducted 45-minute remote sessions via Zoom with children and their parent present. After obtaining parental consent and child assent, we began with a brief demonstration of the tool's features. Children then selected a word to investigate from a set of five words (selected from the from High Frequency Words Project~\cite{loveless2024high} )---\textit{friend}, \textit{definitely}, \textit{favorite}, \textit{because}, \textit{beautiful}---to explore using our spelling tool. During the intervention, children were provided with a pre-written sample where their selected word was intentionally misspelled for their morphological investigation.

Following the intervention, we conducted a retrospective think-aloud protocol, asking children to reflect on the decisions they made while using the tool. Children then completed a 10-item system usability scale questionnaire adapted for children using the Flesch-Kincaid readability scale, using a 5-point Likert scale. Parents observed throughout and provided informal feedback. The children received a small toy as compensation.

\subsection{Results}
Our preliminary evaluation revealed several key findings about the tool's usability and pedagogical potential. Despite the small sample size ($n = 7$), which limits statistical generalization, consistent patterns emerged that provide actionable insights for design iteration.

\subsubsection{Engagement and Perceived Value}
Children expressed enthusiasm for continued use of the tool (median = 4, range = 3--5). More significantly, during the retrospective think-aloud, all seven participants spontaneously discussed how the tool could help them learn words beyond just fixing spelling errors. A fourth grade participant shared, \textit{``I can learn new words with this!''}, referring to the word-matrix mini-game he had to complete while learning about \textit{``because''} and how its base \textit{``cause''} can be used to make many new words like \textit{``causal''} and \textit{``causes''}. Parents echoed this enthusiasm, with one parent observing, \textit{``At school I don't think he gets taught this way.''} Another parent remarked, \textit{``He started with one [misspelling] but walked away with so many more words.''} This aligns with a key goal highlighted by experts in our technical evaluation: utilizing structured word inquiry to facilitate the acquisition of multiple related words from a single investigation.

\begin{figure}
    \centering
    \includegraphics[width=\linewidth]{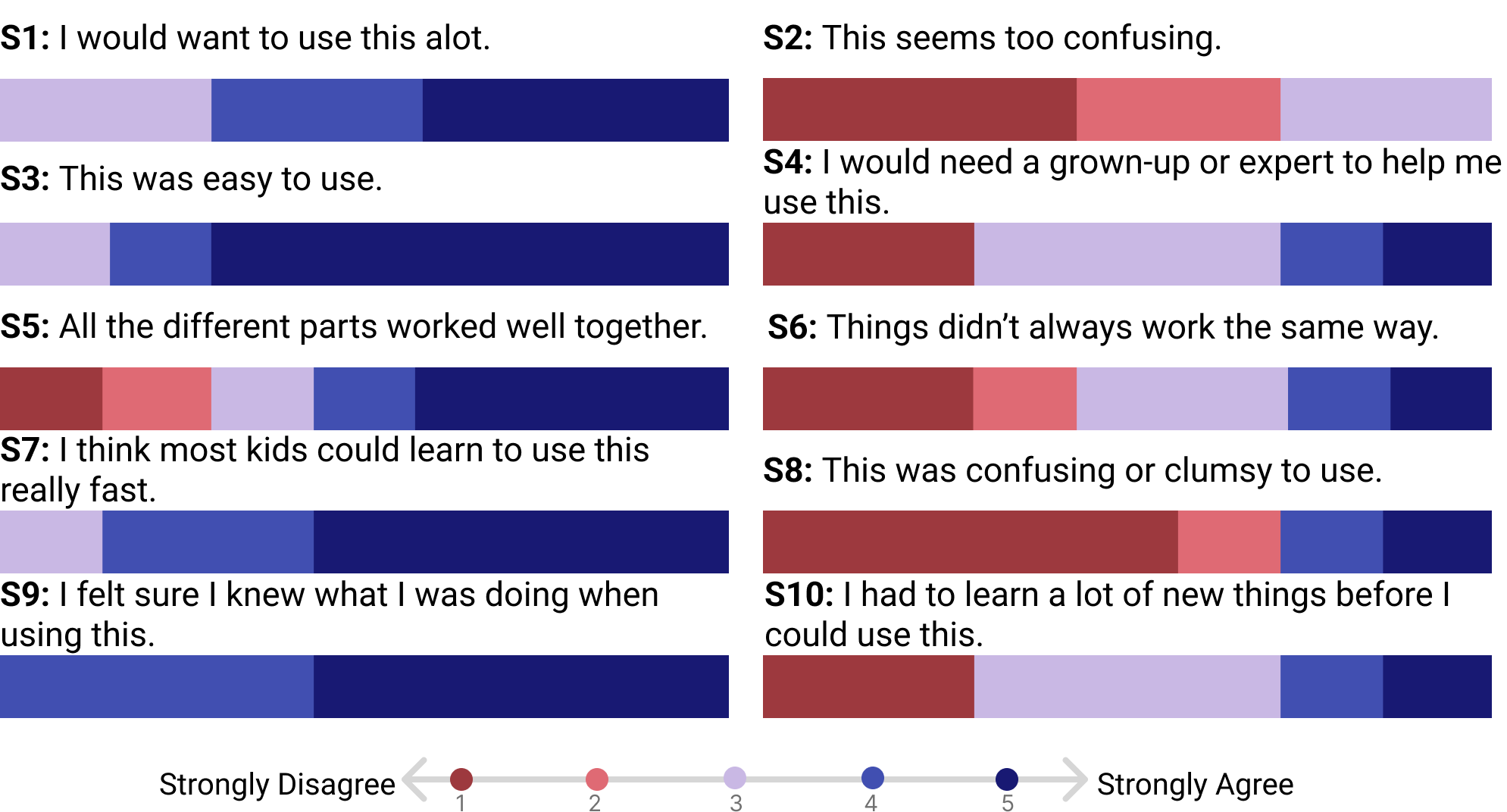}
    \caption{System Usability Survey responses from 7 children. On the left column, higher is better, while on the right column, lower is better.}
    \label{fig:sus}
\end{figure}

\subsubsection{Usability and Learning}

Children responded positively to the tool's usability across multiple dimensions (Figure~\ref{fig:sus}). Most children found the interface easy to use (median = 5, range = 3--5) and felt confident while using it (median = 5, range = 4--5). One fourth-grade participant stated, \textit{``[It was easy because] everything was right there and I didn't have to click too much.''} Children generally felt the tool was not overly complicated (median = 2, range = 1--3), and that other children could learn to use it quickly (median = 3, range = 1--5), suggesting the tool has a small learning curve and can work across age and grade levels. A second-grade participant noted, \textit{``First it was confusing but then I could do it myself.''}

\subsubsection{Discovering Orthographic Structure}

A striking finding emerged regarding participants' prior spelling instruction: none of the children reported having been explicitly taught English orthography through morphological analysis. When they made spelling errors in school, they typically received either direct corrections or phonetic explanations. One participant who has been struggling with spelling recollected, \textit{``My teacher usually tells me to sound the word out.''} All participants unanimously stated that this morphological approach made learning words more enjoyable and meaningful.

\subsubsection{Interface Design}

Children appreciated how the tool guided them through instructions autonomously, making the experience feel more dynamic than filling out a worksheet. Observing the incorrect spelling instance transform gradually into the correct spelling helped them in most instances understand why their hypothesis regarding the spelling was wrong. They liked the different modalities available for input---draggable icons, audio input, and text fields. However, a parent suggested having more animations in the interface to make it more playful and fun for children.

These findings suggest that the tool addresses a gap in current spelling instruction and demonstrates baseline usability for the target age group. While our sample size prevents statistical claims, the consistent positive responses and identification of morphological learning as novel indicate the tool's potential for broader impact.

%% file: 08_discussion.tex
\section{Discussion}

\subsection{Pedagogical Program Synthesis for Personalized and Individualized Learning}

Traditional rule-based systems for instructional support perform well in domains with well-defined constraints and limited solution spaces, such as mathematical problem solving~\cite{siddiqui2024htn}, where errors follow predictable patterns. However, literacy instruction, particularly spelling instruction grounded in morphological, phonological, and etymological analysis, presents different challenges. The cognitive processes expert educators employ when diagnosing spelling errors and constructing instructional responses cannot be adequately captured through explicit rules alone~\cite{fournier2010building}. A single misspelling can involve multiple overlapping dimensions: grapheme-phoneme mappings, morphological boundaries, suffixing conventions, semantic appropriateness, or etymological patterns. The space of pedagogically valid responses is not enumerable in advance, and the optimal instructional sequence depends on nuanced inference about the learner's underlying reasoning. This inference requires flexible interpretation rather than pattern matching against predefined templates.

Pedagogical Program Synthesis addresses this limitation by combining the structured reasoning of domain-specific languages with the interpretive flexibility of large language models. The DSL provides instructional scaffolding, encoding hypothesis templates, evidential features, and compositional primitives that preserve pedagogical validity, while LLMs enable dynamic adaptation to individual learners' contexts. Critically, the DSL also serves as a guardrail against the hallucination risks inherent in generative systems by constraining interventions to valid pedagogical templates and evidence-backed linguistic features. The architecture bounds the space of possible outputs to coherent responses, even when the underlying model produces imperfect completions. This architecture opens possibilities that purely corrective systems cannot achieve. A learner initially seeking to correct a single grapheme error may discover, through synthesized inquiry sequences, the morphological family structure explaining that grapheme choice, the etymological origins preserving it across related words, and the systematic patterns governing similar cases. The learning effect extends far beyond the immediate correction, building metalinguistic awareness that may be transferred to future encounters. From a deployment perspective, the current implementation relies on commercial LLM APIs, which raises legitimate concerns about cost and latency in school settings. However, these constraints are architectural rather than fundamental. Asynchronous background generation and prompt caching can reduce perceived latency. Additionally, future work could explore WebGL-enabled browser-based models for low-cost deployment.

Current classroom instruction rarely provides this form of structured, individualized morphological investigation. \system demonstrates that Pedagogical Program Synthesis can serve as a vessel for structured literacy experiences that do not currently exist at scale, capturing the dynamic processes that expert educators employ while making them accessible within everyday writing contexts. The approach suggests a broader applicability to instructional domains that require flexible and context-sensitive reasoning about learner understanding.

\subsection{Clinical Implications}

\system's architecture is designed to align with the Multi-Tiered Systems of Supports~\cite{cde_mtss} frameworks commonly implemented in schools. While our current evaluation focused on usability and expert validation, the system's modular design supports potential deployment at multiple intervention levels: classroom teachers could integrate it during universal instruction (Tier 1), specialists could use it for targeted small group support (Tier 2), and speech-language pathologists could incorporate it within intensive individualized intervention (Tier 3). Beyond educator-mediated instruction, the system supports independent student practice, facilitating distributed practice across various learning contexts, a condition known to enhance skill generalization beyond clinical settings.

A critical limitation in current writing instruction concerns the temporal misalignment between error production and corrective feedback. Traditional classroom workflows require educators to collect student writing, perform a delayed review, and provide asynchronous feedback. This sequence separates the correction from the motivational and cognitive context in which the errors occurred. \system addresses this temporal gap by providing instruction precisely when learners make spelling errors, capitalizing on the moment of authentic confusion to introduce systematic orthographic investigation. This immediacy transforms errors from failures that require correction into diagnostic opportunities for structured inquiry, aligning the pedagogical intervention with the readiness and intrinsic motivation of the learner.

The system also addresses a persistent resource constraint in specialized literacy support. Speech-language pathologists face significant service demands that limit their capacity to deliver intensive intervention to all students who need support. By automating the synthesis of instructional sequences grounded in pedagogy, \system does not replace clinical judgment, but rather extends the SLP capacity by operationalizing established instructional principles within everyday writing tools where students spend substantial time.

A unique and significant contribution of \system lies in its dual function as both a student intervention and a professional learning resource. It provides a practical, contextualized model for understanding the English orthographic system, emphasizing the critical interrelationship between morphology, etymology, and phonology. Given that many educator training programs fail to provide explicit instruction on the logical and structural principles of English spelling, often leaving the impression that the system is illogical, \system addresses directly this pervasive training gap. By allowing educators to learn and apply these principles alongside their students, it supports continuous professional development and enhances the fidelity of literacy instruction in the school environment.

\subsection{Limitations and Future Work}
\system's current architecture relies extensively on API calls to commercial large language models: Anthropic's Claude Sonnet for hypothesis filtering, trace generation, and program synthesis, and Google's Gemini for speech-to-text and text-to-speech processing. This dependency introduces two practical constraints. First, the cumulative cost of API calls scales linearly with usage, creating potential barriers to widespread deployment in resource-constrained educational settings. Second, these general-purpose models contain vastly more parameters and capabilities than required for the specific tasks \system performs, representing computational overhead that lighter, task-specialized models could avoid. Local models fine-tuned for particular subtasks, such as extracting morphological properties from target words or filtering hypothesis templates based on error patterns, could substantially reduce both latency and cost while maintaining pedagogical quality. However, the primary focus of this work was establishing the viability of Pedagogical Program Synthesis as an instructional paradigm and demonstrating alignment with expert practice. System optimization through custom model training represents an important engineering direction for future work, but was outside the scope of this initial investigation.

Our empirical evaluation, while providing valuable insight into the pedagogical validity and usability of \system, involved modest sample sizes: ten speech-language pathology experts and seven children. The expert evaluation yielded strong evidence of pedagogical soundness, with near-ceiling ratings and high inter-rater agreement on reasoning validity. The preliminary user study with children revealed consistent patterns that suggested good usability and engagement. However, these sample sizes exclude robust quantitative claims about effectiveness or generalizability in diverse populations of learners, grade levels, and ranges of spelling proficiency. In particular, future work should include more students who are not native English speakers and whose native language may have a different grammatical structure, such as Chinese, Russian or Spanish. A larger-scale study would strengthen statistical power and enable a more nuanced analysis of how \system's effectiveness varies between learner characteristics. 

Most critically, our evaluation captured only immediate usability and pedagogical appropriateness, not learning outcomes over time. A longitudinal deployment study tracking children's spelling development, transfer of morphological reasoning to untrained words, and sustained engagement across authentic writing tasks would provide compelling evidence of \system's instructional impact. Such work would also illuminate how repeated exposure to structured word inquiry through \system influences learners' metalinguistic awareness and independent word investigation strategies—outcomes central to the pedagogical philosophy motivating this research.

%% file: 09_conclusion.tex
\section{Conclusion}

This work addressed a fundamental gap in literacy instruction: while speech-language pathologists employ inquiry-based approaches to teach spelling through structured word investigation, everyday writing tools remain limited to detection and autocorrection. We introduced \system (Spelling Inquiry Engine), which brings structured word inquiry into the act of composition through Pedagogical Program Synthesis, a novel approach that operationalizes the inherently dynamic pedagogy of spelling instruction.
Central to our contribution is a domain-specific language that represents SLP instructional moves as composable primitives, enabling the system to synthesize tailored programs in real-time from learner errors and render them as interactive interfaces. Each synthesized program guides learners through brief cycles of hypothesis testing on word meanings, structures, morphological families, origins, and grapheme-phoneme correspondences, transforming spelling errors into opportunities for metalinguistic reasoning alongside correction.
Evaluation with ten speech-language pathology experts demonstrated strong pedagogical validity, with near-ceiling ratings on reasoning appropriateness and high interrater agreement. A preliminary usability assessment with seven children revealed positive engagement and identification of morphological investigation as a novel and meaningful learning experience absent from current classroom instruction.
\system demonstrates that program synthesis can capture expert pedagogical reasoning that purely rule-based systems cannot, offering a vessel for structured literacy experiences at scale. This work suggests broader applicability of Pedagogical Program Synthesis to instructional domains requiring flexible, context-sensitive reasoning about learner understanding, extending beyond spelling to other areas where expert practice must adapt dynamically to individual learner needs.

\section{GenAI Usage Disclosure}
In compliance with the ACM policy on the usage of Generative AI, we disclose that ChatGPT (OpenAI GPT-4/5) and Claude (Sonnet 4-5) was used for text refinement and code scaffolding. The authors maintained full intellectual control over the research design, analysis, interpretation, and writing. All generated content was reviewed and revised to ensure factual accuracy and originality.

\begin{acks}
  We are grateful to the reviewers and our study participants for their time and helpful feedback. This work is supported through the AI Research Institutes program by the National Science Foundation and the Institute of Education Sciences, U.S. Department of Education through Award $\#2229873$ - National AI Institute for Exceptional Education.
\end{acks}